\documentclass[submission,copyright,creativecommons]{eptcs}
\providecommand{\event}{FAVO 2011} 
\usepackage{alltt}

\title{A structured approach to VO reconfigurations through Policies}
\author{Stephan Reiff-Marganiec
\institute{Department of Computer Science\\
University of Leicester\\
Leicester, UK}
\email{srm13@le.ac.uk}
}
\def\titlerunning{VO reconfigurations through Policies}
\def\authorrunning{S. Reiff-Marganiec}
\begin{document}
\maketitle

\begin{abstract}
One of the strength of Virtual Organisations is their ability to dynamically and rapidly adapt in response to changing environmental conditions. Dynamic adaptability has been studied in other system areas as well and system management through policies has crystallized itself as a very prominent solution in system and network administration. However, these areas are often concerned with very low-level technical aspects. Previous work on the \textsc{Appel} policy language has been aimed at dynamically adapting system behaviour to satisfy end-user demands and -- as part of \textsc{StPowla} -- \textsc{Appel} was used to adapt workflow instances at runtime. In this paper we explore how the ideas of \textsc{Appel} and \textsc{StPowla} can be extended from workflows to the wider scope of Virtual Organisations. We will use a Travel Booking VO as example.
\end{abstract}

\section{Introduction}

Over the last decade there were many changes to the business sector, many were driven by a fundamental penetration of IT into business and daily life. Customers nowadays expect businesses to be available 24/7 and to cater for their ever changing demands -- and they can do so because the world has in many ways become more local. Through the internet one can now obtain services and products from around the globe with no additional effort to buying from a local provider. Clearly these changes mean that businesses have to adapt to cater for this demand and remain competitive.   

This might sound very negative, but many opportunities have arisen out of this change as businesses can now provide to a global market, thus having significantly expanded their customer reach. Businesses can also use this connected world to work together effectively, thus being able to jointly offer services which they cannot offer individually.

On the IT side, Business Process Management (BPM) and SOA have become more integrated, providing promising solutions to the design and development of the software systems of the future, as clearly stated in \cite{Kamoun07}: ``The BPM-SOA combination allows services to be
used as reusable components that can be orchestrated to support the needs of dynamic business processes. The combination enables businesses to iteratively design and optimize business processes that are based on services that can be changed quickly, instead of being `hard-wired'. This has the potential to lead to increased agility, more transparency, lower development and maintenance costs and a better alignment between business and IT.''

These changes apply to the business process, where the flexibility to customize a core model to adapt it to various requirements and to accommodate the variability of a business domain are required. In previous work \cite{wesoa07}, we defined StPowla -- to be read like `Saint Paula' -- a Service-Targeted Policy-Oriented WorkfLow Approach. StPowla supports policy-driven business modelling over general Service Oriented Architectures (SOAs). 

However, the demand for flexibility and copying with variability does not stop at the business process -- it extends to the whole organisation. This becomes of fundamental importance when we consider small and medium organisations working together to be competitive in a global market. In some sectors, for example the building trade, people have done this for a long time by pulling in skills from other companies by subcontracting. Virtual Organisations have emerged as a way to capture flexible cooperation. As such a VO presents a loosely bound consortium of organisations that work together to achieve a specific goal for a customer; the consortium will usually disband when no further demand for its service exists. To allow for VOs to be formed quickly, the notion of a VBE (a VO Breeding Environment) provides a structure in which companies are aware of each other and have established relationships on which they can draw when new demands arise. The organisational advantages of VBEs have been discussed in detail in several publications, for example \cite{Afsarmanesh-2005,Camarinha-Matos-2005}. 

Much work in the area of VOs focuses on their use in organisational environments or steps towards frameworks for enabling VOs. another avenue that is pursued, possibly to a lesser extent, is that of modelling VOs to understand and analyze their behaviour and structure more formally. Such approaches include \cite{Bryans06formalmodelling,PROVE11,prove11bryans}, and the work presented here considers modelling of reconfigurations.

In this short exploratory paper we will review \textsc{Appel} and \textsc{StPowla} in Section \ref{StPowla} and briefly recap on the VOML modelling framework for VOs in Section \ref{VOML}. The core of the paper (Section \ref{reconf})will present initial investigations in extensions to the StPowla ideas for VOs -- which is very much work in progress. Sections \ref{relwork} and \ref{concl} will round the paper off by looking at related work and drawing conclusions on our work.

\section{Appel Policies and Workflow Reconfigurations}\label{StPowla}

Policies have been used to describe rules that are used to modify the behaviour of a system at runtime, e.g.~\cite{Lupu-Sloman-1999}. Much effort has been invested in defining policy languages for low-level system administration (such as management of network routers), but also into policy languages for access control. \textsc{Appel}~has been defined with a natural language semantics \cite{Turner-Reiff-etal-2006} as well as a formal semantics based on $\Delta$DSTL \cite{FUNDAMENTAE}. \textsc{Appel} was developed for telecommunications systems, however it is a general language for expressing policies in a variety of application domains. \textsc{Appel} was designed with a separation between the \emph{core} language and its specialization for concrete \emph{domains}. 

In \textsc{Appel} a \emph{policy} consists of a number of \emph{policy rules}, grouped using a number of operators (\textbf{seq}uential, \textbf{par}allel, \textbf{g}uarded and \textbf{u}nguarded choice). A policy rule has the following syntax
\begin{equation}
	{\normalfont [\textbf{appliesTo}\  location]\ [\textbf{when}\  trigger]\  [\textbf{if}\  condition]\  \textbf{do}\  action} \nonumber
\end{equation}
The core language defines the structure of the policies, the details of these parts are defined in specific application domains. Triggers and actions are domain-specific. An atomic condition is either a domain-specific or a more generic (e.g.~time) predicate. This allows the core language to be used for different purposes.

The applicability of a rule is defined on the core language and depends on whether its trigger has occurred and whether its conditions are satisfied. Triggers are caused by external events. Triggers may be combined using \textbf{or}, with the obvious meaning that either is sufficient to apply the rule. Conditions may be negated as well as combined with \textbf{and} and \textbf{or} with the expected meaning. A condition expresses properties of the state and of the trigger parameters. Finally, actions have an effect on the system in which the policies are applied. A few operators (\textbf{and}, \textbf{andthen}, \textbf{or} and \textbf{orelse}) have been defined to create composite actions. In addition policies are `located', that is one can define to which actor or component of a system they apply.

\textsc{StPowla} is concerned with the adaptation of workflows, which are then executed on top of a service oriented architecture. A workflow defines the business process core as the composition of building blocks called \emph{tasks}, similar to \cite{BPMN}. Each task performs a step in the business; policies are used to express finer details of the business process, by defining details of task \emph{executions} as well as workflow adaptations. Policies can be updated dynamically, to adapt the core workflow to the changing needs of the environment. Note that the policies are assumed to be applied at execution time of a workflow and dynamically make changes to an instance of the workflow that is being run.

Policies in \textsc{StPowla} exist in two flavours: refinement policies and reconfiguration policies. The former are concerned with adding extra requirements to be considered when a service is located to execute a task (e.g.~select only services that are based in Europe). The latter
modify the workflow structure, for example by adding and/or deleting tasks. Let us just consider a few examples of policies, at a quite abstract level. Considering a hotel setting, we might have policies like
\vspace{-0.2cm}
\begin{description}
    \item[P1:] Company invoiced customers will not be required to pay a deposit on checking in.
    \item[P2:] In a small hotel, the hotel manager will show VIP customers to their rooms.
\end{description}

P1 would be an example of a policy that changes the workflow: one would normally expect a `pay deposit' task to exist, which would be removed for the instances covered by P1. P2, on the other hand, is a policy that places requirements on a task: if we assume a `show customer to room' task, then this would normally need to be enacted by a hotel employee and the specific one to be chosen for VIP guests is the hotel manager. 

\textsc{Appel} was specialised to the domain of workflows, by making precise the triggers, conditions and actions that are possible in the domain of workflow adaptation \cite{SensoriaBook} and \cite{FACS} are concerned with refinement and reconfiguration policies respectively. 

The defined triggers are based on tasks: one might wish to apply rules when a task is started, completes or fails so triggers are defined as $task\_entry$, $task\_exit$ and $task\_fail$. Actions are either a parametrised refinement action $req(-, -, -)$ or actions to insert and remove tasks ($insert(T1, T2, -)$ and $remove(T)$). The semantics of $req$ is to \emph{find} a service as described by the first and third arguments (specifying service type and SLA constraints), \emph{bind} it, and \emph{invoke} it with the values in the second argument (the invocation parameters). $insert$ inserts the task specified in the first argument after or in parallel to the task in the second argument based on the value in the third argument while $remove$ removes the task. Clearly when adding or removing tasks there might be clean-up actions needed to ensure a well-formed workflow, which are considered in the definition of the semantics of the actions. Also, $req$ might lead to requirements that cannot be fulfilled if no suitable service can be found -- a fact of reality. 

\section{VOML Models}\label{VOML}

The Virtual Organization Modelling Language (VOML) is dedicated to VO development in the context of a VO Breeding Environment. The VOML approach \cite{FAVO09,PROVE11} supports the definition of structural and behavioural models of VBEs and VOs based on three different levels of representation: (1) the definition of the persistent functionalities of the VBE; (2) the definition of the transient functionalities of the VOs that are offered by the VBE at a specific moment in time and (3) the ensemble of components (instances) and connectors that, at that time, deliver the services offered by the VOs present in the business configuration. 

VOML offers several sub languages, addressing different levels of a VO. VO-S (the structural VO modelling language) is concerned with the structural level description of a VO, VO-O (the operational VO modelling language) is more focused on a description that refines the structural model by providing operational details. VO-R (the VO reconfiguration language) presents an alternative approach to describing reconfigurations (we will contrast this in the next section).

In this paper we are mostly concerned with the structural representation of a VO, as it is here that in our opinion the most interesting adaptations can be made. We will here review VO-S briefly.

We define the basic structure of the VO in VO-S. The VO structural model consists of five basic elements: (1) Members, (2) Process, (3) Tasks, (4) VBEResource and (5) Data-Flow. Of these, Members, Tasks and VBEResources are elements that can occur in VBE specifications, too.

\textit{Members} can be Partners (permanent members of the VBE), Associates (transient members of the VBE who have joined temporarily to fulfil demand for a VO) and extEntity (transient members of a VO who are discovered for each VO instance). The \textit{Process} describes the workflow which lists those tasks that contribute directly towards achieving the goals of the VO (this captures the control flow) and \textit{Data-Flow} expresses which data items are expected from the customer and partners and their flow between tasks. \textit{VBEResources} are resources available to all VOs and are provided by the VBE. Finally, \textit{Task} specifications define the competencies required by the VO from its members. Tasks are complex and a more detailed description is available in \cite{PROVE11}. VO-S provides three types of tasks: AtomicTask (tasks to be performed by only one member), ReplicableTask (tasks that can be shared to gain extra capacity) and ComposableTask (tasks that can be shared to address capability issues) to address one of the main reason of VO formation (namely the incapacity of each individual organisation of reacting to a demand).

\section{VO reconfigurations}\label{reconf}

Applications of \textsc{Appel} in the context of communications systems, and indeed in our related work \cite{PROVE11} policy triggers are events occurring through communication or events in the system. For example, we have suggested triggers such as $capability\_deficit()$, which would be raised if the VO (maybe through its coordinator) identifies that there could be the specific problem of a certain capability not being available.  The advantage is that the events are occurring in the domain, however it comes with the disadvantage that new such domain events might be formulated and hence the vocabulary of the policy languages needs to be extended. There is also the question as to implementing the monitoring of the environment and the raising of the triggers as part of a component or actor who gains a very central role in the VO. 

\textsc{StPowla} considers adaptations made to the workflows or tasks in relation to the starting or entering of a task, so one could say that policies are triggered by the execution structure of the workflow. This approach is different in that decisions to restructure are taken just before a task is undertaken or after it has finished (or failed). This means that a very small and stable number of triggers exists. We will explore here how this approach would work for VOs.

Let us consider the specific triggers, actions and conditions that form the domain specific parts of a policy language for VO reconfigurations.

The VO-S model has tasks as central entities, which are then discharged by VO members. It seems sensible to assume that these tasks do in their fundamental nature not differ from tasks in a workflow, and hence we can reuse the three triggers from StPowla. The triggers are summarised in Tab.~\ref{tab:Triggers} 

\begin{table}
	\centering
		\begin{tabular}{ll}
			\hline Trigger         & Description \\ \hline
			$task\_entry()$   & occurs when a new task is started, that is when data and control flows have \\
			                & reached a task and it becomes active \\
			$task\_exit()$    & occurs when a task completes successfully, that is when data and control \\
			                & flows are exiting the task \\
			$task\_failure()$ & occurs when a task fails \\ \hline
		\end{tabular}
	\caption{Triggers}
	\label{tab:Triggers}
\end{table}
  
Note that a tasks will have a hidden boot-strap process where it will be checked whether members are assigned to enact it and where the existence of all capabilities and capacities needed for the task will be evaluated. One could say that this is a default policy for a task, which as actions adds required members and assigns them to the task if there is any shortage. If no suitable members can be found the task will fail.

Policies achieve change through the actions that they propose. In order to understand the full catalogue of actions that we need to offer we need to analyse what aspects of  VO we might wish to change. Recall that \textsc{StPowla} allowed for changes to the workflow as well as the addition of extra requirements for a task. VOs are more complex than workflows, in fact workflows form just one part of a VO specification.

\begin{table}
	\centering
		\begin{tabular}{ll}
			\hline Action & Group \\ \hline
			$add\_task(T1, T2, relation)$ & Workflow -- Control \\
			$delete\_task(T)$ & Workflow -- Control   \\ 
			$provide\_input(I, T)$ & Workflow -- Data \\
			$remove\_input(I, T)$ & Workflow -- Data  \\ 
			$change\_type(T, new_type, args)$ & Task Structure \\
			$add\_member(P)$ & Members \\
			$remove\_member(P)$ & Members \\ 
			$assign\_duty(P, T, args)$ & Duty \\ \hline
			$unassign\_duty(P, T, args)$ & Duty \\ \hline
		\end{tabular}
	\caption{Actions}
	\label{tab:Actions}
\end{table}
 
Table \ref{tab:Actions} shows the actions that will be available to policy authors and hence the opportunities to change the VO. The actions fall into a number of different groups, depending on which entity they affect. We believe that all structural changes that must be made to a VO can be achieved by the above  actions, which we will now describe and analyse in more detail.

The most obvious change to a VO is the addition or removal of a member. To that extend $add\_member()$ and $remove\_member()$ actions are provided. They take a single argument, the member identifier $P$ of the affected member. The semantics of $add\_member()$ is straight forward: a new member will be available in the VO to undertake duties. $remove\_member()$ is more complicated. While the member is removed from the VO and hence will not be available for any duties any more they will have to discharge any activities that they are involved with in active instances of the VO and as a consequence of them being removed the will be automatically unassigned from any duties on tasks. This will possibly lead to a number of tasks that are unassigned to a member or where there is a shortage of capabilities or capacities -- these circumstances will be repaired by the task's default policy. 

$assign\_duty(P, T[, args])$ allows for member $P$ to be added to task $T$. As members could offer a number of capabilities and there is always choice on capacity the assignment operation will allow to specify which capability and what capacity a member will bring to a task. $unassign\_duty(P, T[, args])$ allows to remove responsibilities from a member -- note that as with $remove\_member()$ the member will need to discharge currently active tasks. Either of these two actions will allow to change the involvement of a member by providing a $T$ and $P$ pair that already exists with new values for args -- these will overwrite the existing values. In case of a reduction, again any commitment made to currently active tasks remains. 

Changes to the task structure, allowing for tasks to be changed between \textit{Atomic}, \textit{Replicable }and \textit{Composable }Tasks can be achieved using the $change\_type(T, new\_type, args)$ action.    

We also cater for 4 actions in relation to the workflow of the VO. Two of these are concerned with control flow, while the other two allow to redirect data flow. $add\_task(T1, T2, relation)$ allows to insert a new task $T1$ `next to' task $T2$ -- where `next to' is made precise by the $relation$. The values foreseen for $relation$ are ``parallel'' and ``after'', with the obvious semantics of the control flow reaching the parallel task $T1$ simultaneously with $T2$, whereas the outgoing control flow of $T2$ would be input to $T1$ and $T1$'s output goes to where $T2$'s went before with the after relation. $delete\_task(T)$ deletes a task, making good any control flow breaks caused. Clearly tasks can only be deleted if they are not active. $provide\_input(I, T)$ and $remove\_input(I, T)$ allow to redirect the dataflow, with either providing a data item $I$ to task $T$ or removing it.  

Conditions in policies will usually need to allow checks for certain assignments. We will assume the general checks for time or date that are useful for all policies in all domains. Specifically for VOs we require a few new conditions. These are shown with a brief explanation in Tab.~\ref{tab:Conds}.

\begin{table}
	\centering
		\begin{tabular}{ll}
			\hline Action & Group \\ \hline
			$can\_run(T)$ & allows to check whether all requirements for a task $T$ are full-filled, \\
			             & that is whether all needed members are assigned and sufficient resources\\
									 & are available.\\
			$active(T)$  & allows to check if an instance of task $T$ is executing.\\ 
			$task\_type(T)$ & allows to check the current type of a task $T$.\\
			$has\_capacity(P, c1, c2)$ & allows to check whether member $P$ has free capacity of amount $c2$ \\
			                           & in capability $c1$.\\
			$has\_capability(P, c)$ & short-hand for $has\_capacity(P, c1, 0)$ -- a check whether the member has \\
			                        & a capability.\\ \hline
		\end{tabular}
	\caption{Conditions}
	\label{tab:Conds}
\end{table}

\subsection{Example}

Let us consider a VO for travel arrangements, a quite typical example used  frequently in the literature. \cite{PROVE11} presents a model of such a VO in more detail, however here a brief description will be sufficient. This VO, let us call it VisitUs, offers travel arrangements including flight and hotel bookings. Hotel bookings are provided by a task \texttt{HotelProv}, which for sake of argument is currently fulfilled by one partner with an exclusive contract. VisitUs realises that it could increase business by being able to offer more accommodation. The current hotel provision partner is in agreement to let competitors contribute rooms when they a getting too full. The following policy describes how this could be captured.

\begin{alltt}
\textbf{policy} MoreBeds 
          \textbf{appliesTo} HotelProv
\textbf{when} task\_entry()
\textbf{if not }has\_capacity(Hotel, beds, n)
\textbf{do} change\_type(HotelProv, Replicable, competition) 
	\textbf{andthen} add\_member(newHotel) 
	\textbf{andthen} assign\_duty(newHotel, beds)
\end{alltt}

The policy \texttt{MoreBeds} describes situations where the \texttt{HotelProv} task needs to be adapted if the hotel partner does not have sufficient capacity. In order to allow for other organisations to offer beds, the current partner cannot be exclusive anymore, and hence the task has to become of a replicable type. Once it can be shared, a new hotel partner can be added and be assigned duties against the task. Note that the task can be shared once it is replicable, but here a sharing of a competitive nature has been selected, under which the partner offering the best conditions will get the allocation.

\section{Related Work}\label{relwork}

There has been some effort in modelling virtual organisations, as well as providing flexibility in systems. We would like to identify three specific items here, as we deem them most relevant. However, this is not meant to be an exhaustive review and we are aware that other work that can be seen as relevant exists, which is not mentioned here.

\cite{FAVO-agents} presents a formal set-up to model the structure and responsibilities of a VO in an agent based setting. This work is focused mostly on the creation of VOs which is an aspect that we do not cover here. We do believe that the presented reconfiguration work is orthogonal and and could be applied to the agent based work where the main change would be that the enactment of the actions defined in this paper would be on agents rather than on the VOML model. This could be an interesting angle, as agents are by their very nature much closer to an executable environment than a modelling language. 

VDM has been employed in \cite{Bryans06formalmodelling} to model VOs. This work differs from our approach in that it uses a general purpose modelling language rather than the VO specific one used here. An advantage from their approach is the direct access to verification tools and methodologies, which they employed to analyse properties of VOs. This approach does not provide a direct route to domain experts to model VOs and describe reconfigurations as extensive VDM knowledge is required. Some more recent work \cite{Fitzgerald-2008} proposes the use of animation to make the analysis of the models more accessible -- however it does not address the aspect that we considered here of making the actual modelling more accessible.

While we focused on system adaptation through policies, which has been a long standing and successful approach, it is not the only. Aspect oriented programming allows to isolate concerns and treat them as separate concepts when implementing systems. \cite{Karastoyanova:2009:BAS:1586636.1586901} presented an approach that allows to reconfigure BPEL processes using aspect oriented techniques. They dynamically weave separately defined rules (expressed as aspects) into a BPEL process to adapt its behaviour. This could be a possible alternative to the approach presented here. It would require that the trigger points identified were `hooks' for aspects, which would then be executed at the respective places. However, this approach is far less natural, as it requires additional programmes that will be run instead or in addition to the specified task descriptions while the policy based approach is purely descriptive and furthermore shows simple changes to the existing system which are easier to comprehend.

\section{Discussion and Future Work}\label{concl}

We presented an overview of \textsc{StPowla}, an approach based on the \textsc{Appel} policy language to dynamically adapt workflows. We also reviewed the VOML, the Virtual Organisation Modelling Language, framework briefly, focusing our attention to the structural descriptions for VOs. The aim of the paper was to explore how a \textsc{StPowla}-like approach could be used to adapt virtual organisations. The result of this analysis is an identification of a small and comprehensive set of domain specific triggers, actions and conditions for the \textsc{Appel} policy language that allow to express structural reconfigurations of VOs. This has been exemplified with a Travel Booking VO. We believe that the small policy language presented allows to describe typical structural transformations required for a VO in a natural way. 

There are two more generic aspects worthy of discussion: one is the question whether the added formality gained by modelling is of actual benefit to the VO community and the other is whether the definition of more specific or the use of more generic languages provides a better approach. Considering the former, VOs have been studies from several angles, but like any system it is often useful to have a very precise understanding of their behaviour. one example would be the analysis of scenarios making predictions about the future. For example one might wish to know what happens if a partner were to leave or what would happen if a disaster struck and some services could not be provided anymore. Formal analysis of a model allows to make such predictions, but for that a model is required capturing the desired properties of the VO. This has let to several approaches for modelling VOs as we have shown in the previous section.

The second aspect is a more debatable topic, as the existing studies in the more formal arena show that different groups have taken different views on this. A more specific language offering dedicated concepts for modelling VOs is desirable as it allows `users' of VOs to describe them and to also understand their structure. Our work on VOML falls into this category as it makes concepts from the VO domain first class citizens in the language. Work using general formal modelling languages has a big advantage in that it can usually rely on readily available tools for modelling as well as analyzing said models, however this usually comes at the cost of experts in those languages and tools being required. Of course one could add a further less formal layer, namely that of generic graphical modelling languages such as UML into the discussion. What we believe is that the ideal approach is a combination of all of these techniques. We foresee that for example UML activity diagrams form an ideal vehicle to describe the processes and task dependencies in the workflow of the VO, with VOML (or similar domain specific languages) allowing to express domain specific criteria of VOs. The combination of these could be mapped into a more generic modelling language such as VDM and at that level analysis could be performed. This leaves one very crucial aspect, which has seen more attention in the formal modeling and verification community in the last decade, namely that of mapping the analysis results back into the domain language so that the user can understand the results more directly.

Future work will analyse which properties of a VO can be guaranteed in the light of changes possible through policies -- that is we wish to study how fundamental a VO can be changed through policies and whether that is desirable. The initial feeling is that this is not too desirable, as essentially by using the right policies the purpose of a VO could be changed fundamentally (all tasks could be removed and a completely new set be added). Furthermore, there is of course the questions of how easy it will be to recover through the tasks default policy from member removals or duty dis-allocations. 

A solution to restrict the changes possible, which would not only address this problem, but actually add to the properties one might wish to define on VO models would be to add more policies (also expressed in \textsc{Appel}) at the VBE and VO level. These policies would be constraints rather than reconfiguration instructions, and as such would not have triggers, but rather would specify a set of conditions and have actions that prohibit or undo (or repair) certain reconfigurations made. This will be explored, as it is of interest beyond the scope of the current work presented here. Having policies expressed at different levels also can give rise to a problem referred to as policy conflict: this occurs when policies are simultaneously requesting actions that are incompatible -- however solutions to the policy conflict problem have already been studied and formalised for \textsc{Appel} in \cite{FUNDAMENTAE}. 

\bibliographystyle{eptcs}
\bibliography{favo2011}
\end{document}